\documentclass[a4paper,12pt]{article}

\usepackage{graphicx}

\def\be{\begin{equation}}
\def\ee{\end{equation}}
\def\bea{\begin{eqnarray}}

\def\eea{\end{eqnarray}}

\begin{document}

\title{Thermalization of an Interacting Quantum Field in the
CTP-2PI Next-to-leading-order Large N Scheme}

\author{E. A. Calzetta$\dagger$\thanks{Email: calzetta@df.uba.ar} and
B. L. Hu$\ddagger$\thanks{Email: hub@physics.umd.edu}\\
$\dagger$ Departamento de Fisica,
Facultad de Ciencias Exactas y Naturales\\
Universidad de Buenos Aires- Ciudad Universitaria,\\
1428 Buenos Aires, Argentina\\
$\ddagger$ Department of Physics, University of Maryland,\\
College Park, MD 20742, USA}
\date{May 22, 2002}

\maketitle

\begin{abstract}
In this paper we use an $O(N)$-invariant scalar field of unbroken symmetry
to investigate whether an interacting quantum field at the next-to-leading
order Large $N$ approximation may show signs of thermalization. We develop
the closed time-path (CTP) two-particle irreducible (2PI) effective action in powers of $%
1/N$, retaining up to next to leading order ($O(1)$) terms, and write down
the corresponding (truncated) Schwinger-Dyson equations for its two point
function. We show that in this approximation, the only translation invariant
solutions to the Schwinger - Dyson equations are thermal. This provides a
useful temperature concept without invoking a heat bath. Coupled with the
familiar Kadanoff-Baym approach to quantum kinetic theory our result shows
that at this order of approximation thermalization can occur, at least if
initial conditions are smooth enough that a derivative expansion is valid.
Our analytic result provides support for similar claims in recent literature
based on numerical evidence.
\end{abstract}

\section{Introduction and Summary}

The problem of thermalization in relativistic quantum fields has
drawn much attention over time, both in its own right in our
attempt to understand the origin of macroscopic irreversible
behavior from  microscopic theories, and in the context of
nonequilibrium quantum field processes in the early Universe and
in relativistic heavy ion collisions \cite
{S92,SX93,H01,HK01,HW02,BMSS00,BMSS02}.

The goal of this paper is to investigate whether at the next-to-leading
order (NLO) Large $N$ approximation \cite{CJP74,R74,CHKMPA94,CKMP95} an
interacting quantum field may show signs of thermalization. We consider a $%
O(N)$ invariant scalar field of unbroken symmetry, develop the
closed time-path (CTP) 2-particle irreducible (2PI) \cite{CJT74}
\cite {S61,K64,BM63,CSHY85} effective action (EA) \cite{CH88} in
powers of $1/N$, retaining up to the next-to-leading order
($O(1)$) terms, and write down the corresponding (truncated)
Schwinger-Dyson equations for its two point function. We show
that in this approximation, the only translation invariant
solutions to the Schwinger - Dyson equations are thermal. Thus,
without having it coupled to a heat bath this provides a useful
temperature concept. Together with the familiar Kadanoff-Baym
approach to quantum kinetic theory \cite{CH88,KB62,D84,CHR00},
this result shows that at this order of approximation
thermalization can occur, at least if initial conditions are
smooth enough that a derivative expansion is valid. Our analytic
result provides support for similar claims in recent literature
based on numerical evidence \cite{BC01,B02}. We will not address
the bigger question of whether the thermal solutions are in any
sense an attractor with a nontrivial basin of attraction, as it
is considerably more difficult. (We may look at the theory of
glasses to see just how involved the long time behavior of a
field theory can be \cite{CGLLS01}.)

However, to appreciate this result, we must distinguish between a
`true' $NLO$ approximation (truthful to the scheme) to the full
quantum theory, and the theory which results from solving the \
$NLO$ Schwinger-Dyson equations as if they were exact. By a
`true' $NLO$ approximation we mean that, after deriving the
Schwinger-Dyson equations up to $NLO$ corrections, the solution
to these equations is expanded as well, and terms of higher order
which then result are discarded. In the second -- call it the
`exact'-- procedure, once the equations are written down, a
solution is sought, which will involve terms at all orders in
$1/N$.

To give an example, the situation is similar to the usual textbook
derivation of the running of coupling constants in an interacting
field theory. A renormalization group equation, let us say for a
$\lambda \phi ^{4}$ scalar field theory, is derived within
perturbation theory as

\begin{equation}
\mu \frac{d\lambda }{d\mu }=\epsilon \lambda +a\lambda ^{2}  \label{rg1}
\end{equation}
where $\mu$ is a renormalization parameter. We then solve this equation as
if it were exact and get

\begin{equation}
\lambda \left( \mu \right) =\left( \frac{\mu }{\mu _{0}}\right) ^{\epsilon
}%
\frac{\lambda \left( \mu _{0}\right) }{1-\frac{a\lambda \left( \mu
_{0}\right) }{\epsilon }\left[ \left( \frac{\mu }{\mu _{0}}\right)
^{\epsilon }-1\right] }  \label{rg2}
\end{equation}
Call this the `exact' way. Of course, a Taylor expansion of
$\lambda \left( \mu \right) $ should include all powers of
$\lambda \left( \mu _{0}\right) ,$ but only up to two powers were
considered in Eq. (\ref{rg1}). A true second order approximation
would be restricted to

\begin{equation}
\lambda \left( \mu \right) =\left( \frac{\mu }{\mu _{0}}\right) ^{\epsilon
}\lambda \left( \mu _{0}\right) \left\{ 1+\frac{a\lambda \left( \mu
_{0}\right) }{\epsilon }\left[ \left( \frac{\mu }{\mu _{0}}\right)
^{\epsilon }-1\right] \right\}   \label{rg3}
\end{equation}
Call this the `true' way. We would obtain this same solution if \
$\lambda ^{2}$ in the second term in Eq. (\ref{rg1}) is
approximated by

\begin{equation}
\lambda ^{2}\sim \left[ \left( \frac{\mu }{\mu _{0}}\right) ^{\epsilon
}\lambda \left( \mu _{0}\right) \right] ^{2}
\end{equation}
However, we usually believe that Eq. (\ref{rg2}) is meaningful, although it
oversteps the bounds within which Eq. (\ref{rg1}) was derived, because it
can be shown that it captures the leading terms in the perturbative
expansion (the so-called leading logs), and because in any case we only use
it in the regime in which the coupling constant is small.

In the thermalization problem, this procedure breaks down. If we adhere
strictly to the `true' NLO approximation, then the only thermalization
mechanism left in the theory is binary scattering of on-shell particles.
Since this process conserves particle number, the truncated theory allows
thermalization with non vanishing chemical potential. The `exact' theory,
represented by the full hierarchy, on the other hand, does not conserve
particle number \cite{JY96}, and so the chemical potential must vanish in
true equilibrium states. We must conclude that a ''true'' $NLO$
approximation, in the above sense, fails to describe thermalization.

This problem actually disappears if we solve the $NLO$
Schwinger-Dyson equations `exactly' to all orders in $1/N$ (like
going from Eq. (\ref{rg1}) to (\ref{rg2})). In this procedure the
density of states takes on a Breit-Wigner form, and we have states
with all masses. In particular, now a state with squared momentum
$-p^{2}\geq 9M^{2}$ [we use signature (-,+,+,+) for the Minkowski
metric and $M^{2}$ is the leading order physical mass] may decay
into three on-shell particles whereby particle number is no longer
conserved.

In this way, we see that the theory built on the $NLO$
Schwinger-Dyson equations is able to describe thermalization,
including relaxation of the chemical potential. This relaxation
is a higher than $NLO$ effect, and we must raise the issue of
whether our analysis is still a meaningful approximation to the
full theory. The problem is that in this case there is no analog
of the ''leading log'' concept that validates the running coupling
constant Eq. (\ref{rg2}) over Eq. (\ref{rg3}). On the contrary,
the $NLO$ approximation discards $2\rightarrow 4$ scattering of
on-shell particles, which also violates particle number
conservation. The contribution to the relaxation rate from these
higher order effects is comparable to the decay of off-shell
excitations \cite{CHR00}.

We therefore conclude that, while the next to leading order approximation
describes thermalization, it overestimates the relaxation time, and may not
be a realistic picture of thermalization in actual physical systems.

\subsection{The meaning of thermalization}

Let us begin with a discussion of the exact meaning of thermalization. In
the strictest sense an isolated system depicted by quantum field theory
undergoes unitary evolution and does not thermalize. However, one can still
ask meaningful questions such as whether certain correlation functions may
converge to their thermal forms in some well defined physical limit (weak
thermalization).

As a matter of fact even asking this type of questions is too
ambitious. The Schwinger - Dyson equations for the correlation
functions form an infinite hierarchy. Physical limitations on the
precision of our measurements amounts to truncating the
hierarchy. When certain causal boundary condition such as
Boltzmann's molecular chaos hypothesis is imposed, the truncated
subsystem will show signs of irreversibility and a tendency to
equilibrate. In practical terms we have to deal with a truncated
hierarchy to make the analysis possible. Moreover, nontrivial
(point) field theories are plagued by divergences which can only
be controlled by regularization and renormalization within some
perturbative scheme. Therefore the question of weak
thermalization can only make sense within a chosen approximation
scheme, whether this is a $1/N$ expansion, loop expansion,
expansion in powers of a coupling constant, etc., to allow us to
organize the Schwinger - Dyson equation and  evaluate the
relative weight of different processes.

One must distinguish between two different viewpoints. If one accepts
thermalization as an empirical fact, then there is only the question of
coming up with a formulation which describes this process. For example, one
may assume from the beginning that the relevant dynamics involves only the
longest scales in space and time \cite{AY98,ASY99a,ASY99b,B99}, introduce a
Wigner function as a partial Fourier transform of the propagators (as in the
Kadanoff - Baym approach \cite{CH88,KB62,D84,CHR00}) and deduce a Boltzmann
equation for the dynamics of this Wigner function \cite{BI02,CH00,LM01}. Of
course the resulting model describes thermalization.

However, this is not a proof that the original field theory
thermalizes, even weakly, because there is no clear cut
description of the set of initial conditions for which the
Kadanoff-Baym approach, which depends upon a derivative expansion
of the propagators, is valid. Mrowczynsky has shown that, for
free theories, the only correlators that satisfy a reasonable
almost-invariant condition are those which are exactly
translation invariant \cite{M97}. In other words, the set of
initial conditions which permits thermalization as described by
the Kadanoff-Baym equations may be empty, but for the equilibrium
solution itself. We may say that the Kadanoff - Baym formalism is
useful for studying certain important questions, such as the
determination of the transport coefficients \cite{CHR00},
assuming one already knows on independent grounds that the system
thermalizes. It describes certain important aspects of the
thermalization process, but it does not address the conditions
conducive to it.

On the other hand, if one does not assume thermalization, then one needs an
approximation scheme which avoids the imposition of an arrow of time on the
system by hand.

\subsection{The Large $N$ Approximation Beyond Leading Order}

The number $N$ of replicas of essentially identical fields (like the $N$
scalar fields in an $O(N)$ invariant theory, or the $N^{2}-1$ gauge fields
in a $SU(N)$ invariant non-abelian gauge theory) suggests using $1/N$ as a
natural small parameter, with a well defined physical meaning and that,
unlike coupling constants, is not subjected to renormalization or radiative
corrections. By ordering the perturbative expansion in powers of this small
parameter, several nonperturbative effects (in terms of coupling constants)
may be systematically investigated.

In the case of the\ $O(N)$ invariant theory, in the presence of a nonzero
background field (or an external gravitational or electromagnetic field
interacting with the scalar field) we may distinguish the longitudinal
quantum \ fluctuations in the direction of the background field, in field
space, from the $N-1$ transverse (Goldstone or pion) fluctuations
perpendicular to it. To first order in $1/N$, the longitudinal fluctuations
drop out of the formalism, so we effectively are treating the background
field as classical. Likewise, quantum fluctuations of the external field are
overpowered by the fluctuations of the $N$ scalar field. In this way, the $%
1/N$ framework provides a quantitative measure and concrete meaning to the
semiclassical approximation \cite{HH81}.

To leading order ($LO$), the theory reduces to $N-1$ linear fields with a
time dependent mass, which depends on the background field and on the linear
fields themselves through a gap equation local in time. This depiction of
the dynamics agrees both with the Gaussian approximation for the density
matrix \cite{EJP88,MP89} and with the Hartree approximation \cite{HKMP96}.

The ability of the $1/N$ framework to address the nonperturbative aspects of
quantum field dynamics has motivated a detailed study of the properties of
these systems. In non-equilibrium situations, this formalism has been
applied to the dynamics of symmetry breaking \cite{CKMP95,BVHKP98,BVHS99}
and self-consistent semiclassical cosmological models \cite
{BVHS96,BCV01,RH97a,RH97b,GKLS97}.

The $LO$ $1/N$ theory is Hamiltonian \cite{HKMP96} and time- reversal
invariant. However, it does not thermalize. For example, if we set up
conditions where both the background field and the self-consistent mass are
space-time independent, then the particle numbers for each fluctuation mode
will be conserved. The existence of these conservation laws precludes
thermalization \cite{BW97}.

We note that the failure of the $LO$ approximation to describe
thermalization is indicative of a more general breakdown of the
approximation at later times, where effects of particle
interaction dominate. Both the distribution of energy among the
field modes and the phase relationships (or lack thereof) among
them affect the way quantum fluctuations react on the background
or external fields. Therefore, from physical considerations, one
can say that a theory which does not describe thermalization
becomes unreliable for most other purposes as well \cite{S96}.

This is where the next to leading order ($NLO$) approximation enters. It has
been applied to quantum mechanics \cite{MACDH00,MDC01}, classical field
theory \cite{BW98,ABW00a,ABW00b,BCDM01} and quantum field theory in $1$
space dimension \cite{AB01}, being contrasted both to exact numerical
simulations of these systems, as well as against other approximations
purporting to go beyond $LO$. The $NLO$ has been shown to be an accurate
approximation, even at moderate values of $N.$

The $2PI$ formalism is also suitable for this question because, provided an
auxiliary field is cleverly introduced, the $2PI$ $CTP$ effective action can
be found in closed form at each order in $1/N$  \cite{R74,AABBS02}.

\subsection{This paper}

This paper is organized as follows: In Section II we present our
model, calculate the 2PI CTP EA, and discuss several properties
of the propagators which hold to all orders in $1/N$. In Section
III we discuss translation invariant solutions to the
Schwinger-Dyson equations, still without any explicit
approximations. In Section IV we implement the large $N$
approximation, showing that it is possible to write down a closed
expression for the 2PI CTP EA to next to leading order. In
Section V we show that the only translation invariant solutions
to the NLO Schwinger-Dyson equations are thermal. In Section VI
we discuss the relaxation of the chemical potential. In the final
Section, we ask a central question at the foundations of
statistical mechanics: where does macroscopic irreversibility
arise from microscopic reversible dynamics? We point out the
exact spot where coarse-graining was introduced which leads to the
appearance of thermalization.

\section{The model}

Let us consider a $O(N)$-invariant quantum scalar field $\Phi$, in the limit
$N\rightarrow \infty $. The classical action

\begin{equation}
S=\int d^{d}x\; (-{\frac{1}{2}}) \left\{ \partial _{\mu }\phi_{B}^{\alpha
}\partial ^{\mu }\phi_{B}^{\alpha }+ M_{B}^{2}\phi_{B}^{\alpha
}\phi_{B}^{\alpha }+ \frac{\lambda _{B}}{4N}\left( \phi_{B}^{\alpha
}\phi_{B}^{\alpha }\right) ^{2}\right\}
\end{equation}
where $\phi_{B}^{\alpha }$, $M_{B}^{2}$ and $\lambda _{B}$ are the bare wave
function, mass parameter and coupling constant, soon to be renormalized, and
the dimension $d=4-\varepsilon $. We introduce the bare wave function
renormalization $Z_{B}$ by rescaling $\phi_B ^{\alpha }=\sqrt{\left(
N/Z_{B}\right) }\phi ^{\alpha }$

\begin{equation}
S=\frac{N}{Z_{B}}\int d^{d}x\; (-{\frac{1}{2}})\left\{ \partial _{\mu }\phi
^{\alpha }\partial ^{\mu }\phi ^{\alpha }+ M_{B}^{2}\phi ^{\alpha }\phi
^{\alpha }+ \frac{\lambda _{B}}{4Z_{B}}\left( \phi ^{\alpha }\phi ^{\alpha
}\right) ^{2}\right\}
\end{equation}
Since later on we shall discuss in detail the conservation laws of the exact
and approximated dynamics, let us observe that this theory conserves,
besides energy - momentum, a number of Noether charges associated with the
global $O(N)$ symmetry. Concretely, if the infinitesimal $O(N)$
transformation reads $\phi ^{\alpha }\rightarrow \phi ^{\alpha }+\varepsilon
_{A}T_{\quad \beta }^{A\alpha }\phi ^{\beta }$ then the Noether charges are

\begin{equation}
Q^{A}=\frac{N}{Z_{B}}\int d^{d-1}x\;T_{\alpha \beta }^{A}\phi ^{\beta
}\dot{%
\phi}^{\alpha }
\end{equation}
To investigate thermalization in total generality, we should allow for a
Lagrange multiplier for each of these charges. Here we assume that all of
these vanish, as well as the mean value of the charges themselves. Also,
since there is no particle current the field is its own antiparticle, and
the chemical potential must be zero in equilibrium. Discarding a constant
term, we may rewrite the classical action as

\begin{equation}
S=\frac{N}{Z_{B}}\int d^{d}x\; (-{\frac{1}{2}})\left\{ \partial _{\mu }\phi
^{\alpha }\partial ^{\mu }\phi ^{\alpha }+\left[ \sqrt{\frac{Z_{B}}{\lambda
_{B}}}M_{B}^{2}+\sqrt{\frac{\lambda _{B}}{Z_{B}}}\frac{\phi ^{\alpha }\phi
^{\alpha }}{2}\right] ^{2}\right\}
\end{equation}

To set up the $1/N$ resummation scheme, it is customary to introduce the
auxiliary field $\chi $ writing

\begin{equation}
S=\frac{N}{Z_{B}}\int d^{d}x\; (-{\frac{1}{2}}) \left\{ \partial _{\mu }\phi
^{\alpha }\partial ^{\mu }\phi ^{\alpha }+ \left[ \sqrt{\frac{Z_{B}}{\lambda
_{B}}}M_{B}^{2}+\sqrt{\frac{\lambda _{B}}{Z_{B}}}\frac{\phi ^{\alpha }\phi
^{\alpha }}{2}\right] ^{2}-{\frac{1}{2}} \left[ \sqrt{\frac{Z_{B}}{\lambda
_{B}}}\left( M_{B}^{2}-\chi \right) +\sqrt{\frac{\lambda
_{B}}{Z_{B}}}\frac{%
\phi ^{\alpha }\phi ^{\alpha }}{2}\right] ^{2}\right\}
\end{equation}
whence

\begin{equation}
S=\frac{N}{Z_{B}}\int d^{d}x\;\left\{ -{\frac{1}{2}} \partial _{\mu }\phi
^{\alpha }\partial ^{\mu }\phi ^{\alpha }-\chi \left[ \frac{Z_{B}}{\lambda
_{B}}M_{B}^{2}+\frac{\phi ^{\alpha }\phi ^{\alpha }}{2}\right]
+\frac{Z_{B}}{%
2\lambda _{B}}\chi ^{2}\right\}.  \label{classact}
\end{equation}
From now on, we consider $\chi $ and $\phi ^{\alpha }$ as
fundamental fields on equal footing. We assume a background field
decomposition $\phi \equiv \bar \phi + \varphi$ and that the
background field $\bar \phi$ is identically zero (because of the
$O(N)$ symmetry, the symmetric point must be a solution of the
equations of motion) so we can focus on the dynamics of the
fluctuation fields $\varphi$. We also split the auxiliary field
$\chi $ field into a background $\bar{\chi}$ and a fluctuation
$\tilde \chi $, $\chi =\bar{\chi}+\tilde \chi .$ The action
becomes

\begin{equation}
S=S_{back}+S_{lin}+S_{quad}+S_{cub}
\end{equation}
$S_{back}$ is just the classical action evaluated at $\varphi ^{\alpha }=0,$
$\chi =\bar{\chi}$

\begin{equation}
S_{back}=\frac{N}{\lambda _{B}}\int d^{d}x\;\left\{ \frac{1}{2}\bar{\chi}%
^{2}-M_{B}^{2}\bar{\chi}\right\}
\end{equation}
$S_{lin}$ contains terms linear on $\tilde \chi $ and can be set to zero by
a choice of the background field $\bar \chi$.

\begin{equation}
S_{lin}=\frac{N}{\lambda _{B}}\int d^{d}x\;\left\{ \bar{\chi}%
-M_{B}^{2}\right\} \tilde \chi
\end{equation}
$S_{quad}$ contains the quadratic terms and yields the tree - level inverse
propagators

\begin{equation}
S_{quad}=\frac{N}{Z_{B}}\int d^{d}x\;\left\{ \frac{-1}{2}\partial _{\mu
}\varphi ^{\alpha }\partial ^{\mu }\varphi ^{\alpha }-\frac{\bar{\chi}}{2}%
\varphi ^{\alpha }\varphi ^{\alpha }+\frac{Z_{B}}{2\lambda _{B}}\tilde \chi
^{2}\right\}
\end{equation}
Finally $S_{cub}$ contains the bare vertex

\begin{equation}
S_{cub}=\left( \frac{-N}{2Z_{B}}\right) \int d^{d}x\;\left\{ \tilde \chi
\varphi ^{\alpha }\varphi ^{\alpha }\right\}
\end{equation}

To write the $2PI$ $CTP$ $EA$ we double the degrees of freedom,
incorporating a branch label $a=1,2$. We also introduce propagators $%
G^{\alpha a,\beta b}$ for the path ordered expectation values

\begin{equation}
G^{\alpha a,\beta b}\left( x,y\right) =\left\langle \varphi ^{\alpha
a}\left( x\right) \varphi ^{\beta b}\left( y\right) \right\rangle
\label{eq12}
\end{equation}
and $F^{ab}$ for

\begin{equation}
F ^{ab}\left( x,y\right) =\left\langle \tilde \chi ^{a}\left( x\right)
\tilde \chi ^{b}\left( y\right) \right\rangle
\end{equation}
Because of symmetry, it is not necessary to introduce a mixed propagator, $%
\left\langle \tilde \chi ^{a}\left( x\right) \varphi ^{\beta b}\left(
y\right) \right\rangle \equiv 0$. The $2PI$ $CTP$ $EA$ reads

\begin{eqnarray}
\Gamma &=&S_{back}\left[ \bar{\chi}^{1}\right] -S_{back}\left[
\bar{\chi}^{2}%
\right]  \nonumber \\
&&+\frac{1}{2}\int d^{d}xd^{d}y\;\left\{ D_{\alpha a,\beta b}(x,y)G^{\alpha
a,\beta b}(x,y)+\frac{N}{\lambda _{B}}c_{ab}\delta (x,y)F ^{ab}\left(
x,y\right) \right\}  \nonumber \\
&&-\frac{i\hbar }{2}\left[ Tr\;\ln G+Tr\;\ln F \right] +\Gamma _{Q}
\end{eqnarray}
where $c_{11}=-c_{22}=1$, $c_{12}=c_{21}=0$,

\begin{equation}
D_{\alpha a,\beta b}(x,y)=\frac{N}{Z_{B}}\delta _{\alpha \beta }\left[
c_{ab}\partial^{2}_{x}-c_{abc}\bar{\chi}^{c}\right] \delta (x,y)
\end{equation}
$c_{abc}=1$ when all entries are $1$, $c_{abc}=-1$ when all entries are $2$,
and $c_{abc}=0$ otherwise. $\Gamma _{Q}$ is the sum of all $2PI$ vacuum
bubbles with cubic vertices from $S_{cub}$ and propagators $G^{\alpha
a,\beta b}$ and $F ^{ab}\left( x,y\right).$ Observe that $\Gamma _{Q}$ is
independent of $\bar{\chi}^{c}$.

Taking variations of the $2PI$ $CTP$ $EA$ and identifying $\bar{\chi}^{1}=%
\bar{\chi}^{2}=\bar{\chi}$, we find the equations of motion

\begin{equation}
\frac{N}{2Z_{B}}\delta _{\alpha \beta }c_{ab}D\left( x,y\right) -\frac{%
i\hbar }{2}\left[ G^{-1}\right] _{\alpha a,\beta b}\left( x,y\right)
+\frac{1%
}{2}\Pi _{\alpha a,\beta b}\left( x,y\right) =0
\end{equation}

\begin{equation}
\frac{N}{2\lambda _{B}}c_{ab}\delta (x,y)-\frac{i\hbar }{2}\left[ F ^{-1}%
\right] _{ab}\left( x,y\right) +\frac{1}{2}\Pi _{ab}\left( x,y\right) =0
\end{equation}

\begin{equation}
\frac{N}{\lambda _{B}}\left\{ \bar{\chi}\left( x\right) -M_{B}^{2}\right\}
-%
\frac{N}{2Z_{B}}\delta _{\alpha \beta }G^{\alpha 1,\beta 1}(x,x)=0
\end{equation}
where $D\left( x,y\right) =\left[ \partial^{2}_{x}-\bar{\chi}\left( x\right)
\right] \delta (x,y)$,

\begin{equation}
\Pi _{\alpha a,\beta b}\left( x,y\right) =2\frac{\delta \Gamma _{Q}}{\delta
G^{\alpha a,\beta b}(x,y)};\qquad \Pi _{ab}\left( x,y\right) =2\frac{\delta
\Gamma _{Q}}{\delta F ^{ab}(x,y)}
\end{equation}
We shall seek a solution with the structure

\begin{equation}
G^{\alpha a,\beta b}(x,y)=\frac{\hbar }{N}\delta _{\alpha \beta }G^{ab}(x,y)
\label{eq20}
\end{equation}
which is consistent with vanishing Noether charges. Then it is convenient to
write

\begin{equation}
F ^{ab}(x,y)=\frac{\hbar }{N}H^{ab}\left( x,y\right) ;\qquad \Pi _{\alpha
a,\beta b}\left( x,y\right) =\delta _{\alpha \beta }P_{ab}\left( x,y\right)
;\qquad \Pi _{ab}\left( x,y\right) =NQ_{ab}\left( x,y\right)
\end{equation}
The equations become

\begin{equation}
\frac{1}{Z_{B}}c_{ab}D\left( x,y\right) -i\left[ G^{-1}\right] _{ab}\left(
x,y\right) +\frac{1}{N}P_{ab}\left( x,y\right) =0  \label{schdy}
\end{equation}

\begin{equation}
\frac{1}{\lambda _{B}}c_{ab}\delta (x,y)-i\left[ H^{-1}\right] _{ab}\left(
x,y\right) +Q_{ab}\left( x,y\right) =0  \label{eq23}
\end{equation}

\begin{equation}
\frac{1}{\lambda _{B}}\left\{ \bar{\chi}\left( x\right) -M_{B}^{2}\right\}
-%
\frac{\hbar }{2Z_{B}}G^{11}(x,x)=0  \label{eq24}
\end{equation}
Observe that

\begin{equation}
P_{ab}\left( x,y\right) =\frac{2}{\hbar }\frac{\delta \Gamma _{Q}}{\delta
G^{ab}(x,y)};\qquad Q_{ab}\left( x,y\right) =\frac{2}{\hbar }\frac{\delta
\Gamma _{Q}}{\delta H^{ab}(x,y)}
\end{equation}
These are the exact equations we must solve. The succesive $1/N$
approximations amount to different constitutive relations expressing
$P_{ab}$
and $Q_{ab}$ in terms of the propagators.

\subsection{The retarded propagator}

For later use, we want an equation for the retarded propagator $%
G_{ret}=i\left( G^{11}-G^{12}\right) =i\left( G^{21}-G^{22}\right) $.
Rewrite the Schwinger - Dyson equation above as

\begin{equation}
\frac{1}{Z_{B}}DG^{ac}\left( x,y\right) +\frac{1}{N}\int
d^{d}z\;P_{\;b}^{a}\left( x,z\right) G^{bc}\left( z,y\right) =ic^{ac}\delta
(x,y)  \label{26}
\end{equation}
Subtracting the $(11)$ from the $(12)$ components in the above equation, we
obtain

\begin{equation}
\frac{1}{Z_{B}}DG_{ret}\left( x,y\right) +\frac{1}{N}\int
d^{d}z\;P_{ret}\left( x,z\right) G_{ret}\left( z,y\right) =\left( -1\right)
\delta (x,y)  \label{gret}
\end{equation}
where $P_{ret}=P_{11}+P_{12}.$

\subsection{Some nonperturbative identities}

We note some non-perturbative properties of the self-energy $P_{ab},$ as
follows. From the identity

\begin{equation}
\frac{\delta ^{2}\Gamma _{1PI}}{\delta \varphi ^{\alpha a}\delta \varphi
^{\beta b}}=i\hbar \left[ G^{-1}\right] _{\alpha a,\beta b}
\end{equation}
relating the inverse propagator to the $CTP$ $1PI$ $EA$ $\Gamma _{1PI}$, the
inverse propagators may be read off the Schwinger - Dyson equations, and $%
\Gamma _{1PI}$ may be written as

\begin{eqnarray}
\Gamma _{1PI} &=&\frac{N}{2}\int d^{d}xd^{d}y\;\left\{ \left[ \varphi
^{\alpha 1}-\varphi ^{\alpha 2}\right] \left( x\right) \left[
\frac{1}{Z_{B}}%
D\left( x,y\right) +\mathbf{P}\left( x,y\right) \right] \left[ \varphi
^{\alpha 1}+\varphi ^{\alpha 2}\right] \left( y\right) \right. \\
&&\left. +i\left[ \varphi ^{\alpha 1}-\varphi ^{\alpha 2}\right] \left(
x\right) \mathbf{N}\left( x,y\right) \left[ \varphi ^{\alpha 1}-\varphi
^{\alpha 2}\right] \left( y\right) \right\}
\end{eqnarray}
where $\mathbf{P}$ is causal and $\mathbf{N}$ is even, and both are real.
Computing the derivatives we get

\begin{equation}
\frac{1}{N}P_{11}=\mathbf{P}_{even}+i\mathbf{N}  \label{p11}
\end{equation}

\begin{equation}
\frac{1}{N}P_{12}=\mathbf{P}_{odd}-i\mathbf{N}  \label{p12}
\end{equation}

\begin{equation}
\frac{1}{N}P_{21}=-\mathbf{P}_{odd}-i\mathbf{N}  \label{p21}
\end{equation}

\begin{equation}
\frac{1}{N}P_{22}=-\mathbf{P}_{even}+i\mathbf{N}  \label{p22}
\end{equation}
where

\begin{equation}
\mathbf{P}_{even}\left( x,y\right) =\frac{1}{2}\left[ \mathbf{P}\left(
x,y\right) +\mathbf{P}\left( y,x\right) \right] ;\qquad \mathbf{P}%
_{odd}\left( x,y\right) =\frac{1}{2}\left[ \mathbf{P}\left( x,y\right) -%
\mathbf{P}\left( y,x\right) \right]
\end{equation}
Observe that $P_{ret}=N\mathbf{P}\left( x,y\right) $.

\section{Translation-invariant solutions}

Translation invariant solutions are functions only of the relative variable
$%
x-y$ and may be Fourier transformed

\begin{equation}
G^{ab}\left( x-y\right) =\int \frac{d^{d}p}{\left( 2\pi \right) ^{d}}%
\;e^{ip\left( x-y\right) }\;G^{ab}\left( p\right)  \label{fourier}
\end{equation}
The Fourier transform of an even (odd) kernel is an even (odd) function of
$%
p $. If a kernel is real, the real (imaginary) part of its Fourier transform
is even (odd). Vice versa, if a kernel is imaginary, then the real
(imaginary) part of its transform is odd (even).

It follows that, since $\mathbf{P}_{even}\left( x-y\right) $ is real and
even, $\mathbf{P}_{even}\left( p\right) $ is also real and even, while since
$\mathbf{P}_{odd}\left( x-y\right) $ is real and odd, $\mathbf{P}%
_{odd}\left( p\right) $ is odd and imaginary. We may write

\begin{equation}
\mathbf{P}_{odd}\left( p\right) =i\pi \gamma \left( p\right)
\;\mathrm{sign}%
\left( p^{0}\right)  \label{gamma}
\end{equation}
therefore $\gamma \left( p\right) $ is real and even.

\subsection{The density of states}

Let us introduce the density of states $\Delta \left( p\right) $ out of the
Fourier transform of the Jordan propagator $G=G^{21}-G^{12}$

\begin{equation}
G\left( p\right) =2\pi \Delta \left( p\right) \;\mathrm{sign}\left(
p^{0}\right)
\end{equation}
The Jordan and retarded propagators are related through $G\left( p\right)
=2\;\mathrm{Im}G_{ret}\left( p\right) .$ From Eq. (\ref{gret})

\begin{equation}
G_{ret}=\left( -1\right) \left[ \frac{1}{Z_{B}}D\left( p\right)
+\frac{1}{N}%
\;P_{ret}\left( p\right) \right] ^{-1}
\end{equation}
Because of the retarded boundary conditions on $G_{ret}$, it is understood
that $p^{0}$ is replaced by $p^{0}+i\varepsilon $, $\varepsilon \rightarrow
0 $. We must distinguish two cases. As we shall see below, in the $LO$
approximation, $Z_{B}=1$ and $P_{ret}\left( p\right) =0.$ In this case we
have the explicit expression for $G_{ret}$

\begin{equation}
G_{ret}\left( p\right) =\frac{1}{-\left( p^{0}+i\varepsilon \right) ^{2}+%
\vec{p}^{2}+\bar{\chi}}\qquad (LO)  \label{logret}
\end{equation}
Therefore

\begin{equation}
G\left( p\right) =2\;\mathrm{Im}G_{ret}\left( p\right) =2\pi
\;\mathrm{sign}%
\left( p^{0}\right) \;\delta \left( p^{2}+\bar{\chi}\right) \qquad (LO)
\label{lojordan}
\end{equation}
and

\begin{equation}
\Delta \left( p\right) =\delta \left( p^{2}+\bar{\chi}\right) \qquad (LO)
\label{lodos}
\end{equation}
In all higher approximations, $P_{ret}\left( p\right) \neq 0.$ We get

\begin{equation}
G\left( p\right) =2\left| G_{ret}\left( p\right) \right| ^{2}\mathrm{Im}
\mathbf{P}_{odd}\left( p\right) \qquad (NLO\;\mathrm{and\;higher})
\end{equation}
that is

\begin{equation}
\Delta \left( p\right) =\left| G_{ret}\left( p\right) \right| ^{2}\gamma
\left( p\right) \qquad (NLO\;\mathrm{and\;higher})  \label{gamma2delta}
\end{equation}

\subsection{The distribution function and the fluctuation-dissipation
relation}

Now consider the ($12)$ component of Eq. (\ref{26})

\begin{equation}
\frac{1}{Z_{B}}D\left( p\right) G^{12}\left( p\right) +\frac{1}{N}\;\left[
P_{11}\left( p\right) G^{12}\left( p\right) +P_{12}\left( p\right)
G^{22}\left( p\right) \right] =0
\end{equation}
Introducing the advanced propagator $G_{adv}=\left( -i\right) \left(
G^{22}-G^{12}\right) $ ($G_{adv}\left( p\right) =\left[ G_{ret}\left(
p\right) \right] ^{\ast }$) we may rewrite this as

\begin{equation}
\left[ G_{ret}\left( p\right) \right] ^{-1}=\frac{i}{N}P_{12}\left( p\right)
G_{adv}\left( p\right) \Rightarrow G^{12}\left( p\right) =\frac{i}{N}%
P_{12}\left( p\right) \left| G_{ret}\left( p\right) \right| ^{2}
\end{equation}
and transform this into

\begin{equation}
G^{12}\left( p\right) =\left[ -\pi \gamma \left( p\right) \;\mathrm{sign}%
\left( p^{0}\right) +\mathbf{N}\left( p\right) \right] \frac{\Delta \left(
p\right) }{\gamma \left( p\right) }
\end{equation}
In other words, for a translation invariant solution we must have

\begin{equation}
G^{12}\left( p\right) =2\pi \;F^{12}\left( p\right) \Delta \left( p\right)
\end{equation}
where (recall that $\mathrm{sign}\left( p^{0}\right) =1-2\theta \left(
-p^{0}\right) $)

\begin{equation}
F^{12}\left( p\right) =\theta \left( -p^{0}\right) +f\left( p\right)
\end{equation}
Comparing both expressions for $G^{12}$ we get

\begin{equation}
f\left( p\right) =\frac{1}{2}\left[ \frac{\mathbf{N}\left( p\right) }{\pi
\gamma \left( p\right) }-1\right]  \label{pdf}
\end{equation}
It is more common to write this as

\begin{equation}
\mathbf{N}\left( p\right) =\pi \gamma \left( p\right) \left[ 1+2f\left(
p\right) \right]
\end{equation}
whereby we recognize the fluctuation - dissipation relation.

Given the Jordan and negative frequency propagators, it is easy to find all
the others. In particular, the positive frequency propagator $G^{21}\left(
p\right) =G+G^{12}=2\pi \;F^{21}\left( p\right) \Delta \left( p\right) ,$
with $F^{21}\left( p\right) =\theta \left( p^{0}\right) +f\left( p\right) .$

\subsection{A necessary condition for translation invariant solutions}

The expression for $f\left( p\right) $ above (eq. (\ref{pdf})) is equivalent
to the identity

\begin{equation}
P_{12}\left( p\right) G^{21}\left( p\right) -P_{21}\left( p\right)
G^{12}\left( p\right) =0  \label{neccond}
\end{equation}
Indeed, from eqs. (\ref{p12}), (\ref{p21}), (\ref{gamma}) and (\ref{pdf}) we
get

\begin{equation}
P_{12}\left( p\right) G^{21}\left( p\right) -P_{21}\left( p\right)
G^{12}\left( p\right) =P_{12}\left( p\right) \theta \left( p^{0}\right)
-P_{21}\left( p\right) \theta \left( -p^{0}\right) +\left[ P_{12}\left(
p\right) -P_{21}\left( p\right) \right] f\left( p\right)  \label{aux}
\end{equation}
but also

\begin{equation}
P_{12}\left( p\right) -P_{21}\left( p\right) =2N\mathbf{P}_{odd}\left(
p\right) =2iN\pi \gamma \left( p\right) \;\mathrm{sign}\left( p^{0}\right)
\end{equation}

\begin{eqnarray}
P_{12}\left( p\right) \theta \left( p^{0}\right) -P_{21}\left( p\right)
\theta \left( -p^{0}\right) &=&N\left[ \mathbf{P}_{odd}\left( p\right) -i%
\mathbf{N}\left( p\right) \;\mathrm{sign}\left( p^{0}\right) \right]
\nonumber \\
&=&iN\pi \gamma \left( p\right) \left[ 1-\frac{\mathbf{N}\left( p\right) }{%
\pi \gamma \left( p\right) }\right] \;\mathrm{sign}\left( p^{0}\right)
\nonumber \\
&=&-2iN\pi \gamma \left( p\right) f\left( p\right) \;\mathrm{sign}\left(
p^{0}\right)
\end{eqnarray}
Substituting these identities in Eq. (\ref{aux}) we get (\ref{neccond}),
which is therefore a necessary condition for translation invariant
propagators.

\section{The large $N$ approximation}

So far we have shown that translation invariant solutions are defined by the
density of states $\Delta \left( p\right) $ and the distribution function $%
f\left( p\right) $. To show that they correspond to thermal propagators, we
must show that $f\left( p\right) $ is necessarily of the form of a Bose -
Einstein distribution

\begin{equation}
f\left( p\right) =\left[ e^{\left| \beta p\right| }-1\right] ^{-1}\label{eqpdf}
\end{equation}
To do this, we need explicit expressions for the $P$ kernels, which we can
only find perturbatively.

We shall adopt the large $N$ approximation, which consists of taking the
limit of $N\rightarrow \infty $ in $\Gamma _{Q}$ and retaining only terms
scaling like $N$ ($LO$), $1$ ($NLO$), $N^{-1}$ ($NNLO$), etc. The key
observation is that in any given Feynman graph each vertex contributes a
power of $N,$ each internal line a power of $N^{-1}$, and each trace over
group indices another power of $N.$ We have both $G$ and $H$ internal lines,
but the $G$ lines only appear in closed loops. On each loop, the number of
vertices equals the number of $G$ lines, so there only remains one power of
$%
N$ from the single trace over group labels. Therefore the overall power of
the graph is the number of $G$ loops minus the number of $H$ lines. Now,
since we only consider $2PI$ graphs, there is a minimun number of $H$ lines
for a given number of $G$ loops. For example, if there are two $G$ loops,
they must be connected by no less than $3$ $H$ lines, and so this graph
cannot be higher than $NNLO.$ A graph with $3$ $G$ loops can not have less
than $5$ $H$ lines, and so on.

\subsection{The leading order approximation}

We conclude that $\Gamma _{Q}$ vanishes at $LO,$ and therefore $%
P_{ab}=Q_{ab}=0.$ Under the \textit{ansatz }$Z_{B}=1$, the equations we need
to solve become

\begin{equation}
c_{ab}D\left( p\right) -i\left[ G^{-1}\right] _{ab}\left( p\right) =0
\label{losd}
\end{equation}

\begin{equation}
\frac{1}{\lambda _{B}}\left\{ \bar{\chi}-M_{B}^{2}\right\} -\frac{\hbar
}{2}%
\int \frac{d^{d}p}{\left( 2\pi \right) ^{d}}G^{11}(p)=0  \label{logap}
\end{equation}
We disregard the auxiliary field propagator $H^{ab}$, since to this order it
is decoupled from the background auxiliary field and the other propagators.

For the retarded propagator, we have the expression Eq. (\ref{logret}),
leading to Eq. (\ref{lojordan}) for the Jordan propagator and (\ref{lodos})
for the density of states. Write Eq. (\ref{losd}) as

\begin{equation}
D\left( p\right) G^{ab}\left( p\right) =ic^{ab}
\end{equation}
Setting $a=2,$ $b=1$, we see that $G^{21}$ can only be nonzero at the zeroes
of $D\left( p\right) =-p^{2}-\bar{\chi},$ so we still can write $%
G^{21}\left( p\right) =2\pi \;F^{21}\left( p\right) \Delta \left( p\right)
.$
Also $G^{12}(p)=G^{21}(-p),$ and $G^{21}(p)-G^{12}(p)=G\left( p\right) ,$ so
from Eq. (\ref{lojordan}) we conclude that $F^{21}\left( p\right)
-F^{21}\left( -p\right) =\mathrm{sign}\left( p^{0}\right) .$ Therefore we
may write $F^{21}\left( p\right) =\theta \left( p^{0}\right) +f\left(
p\right) ,$ with $f\left( p\right) $ a real and even function. From these
results, we may write all propagators in terms of the distribution function
$%
f\left( p\right) .$ In particular, the Feynman propagator becomes

\begin{eqnarray}
G^{11} &=&\left( -i\right) G_{ret}+G^{12}  \nonumber \\
&=&\frac{\left( -i\right) }{-\left( p^{0}+i\varepsilon \right) ^{2}+\vec{p}%
^{2}+\bar{\chi}}+2\pi \left[ \theta \left( -p^{0}\right) +f\left( p\right) %
\right] \delta \left( p^{2}+\bar{\chi}\right) \;  \nonumber \\
&=&\frac{\left( -i\right) }{-p^{0}{}^{2}+\vec{p}^{2}+\bar{\chi}-i\varepsilon
}+2\pi \;f\left( p\right) \;\delta \left( p^{2}+\bar{\chi}\right)
\label{logf}
\end{eqnarray}
So, assuming $d=4-\epsilon $ dimensions, we may evaluate

\begin{equation}
\int \frac{d^{d}p}{\left( 2\pi \right) ^{d}}G^{11}(p)=\mu ^{-\epsilon
}\left[
M_{V}^{2}+M_{f}^{2}\right]
\end{equation}
where $\mu $ is some (so far) arbitrary renormalization scale,

\begin{equation}
M_{V}^{2}=\frac{\left( -\mu ^{2}\right) }{2\pi \epsilon }\frac{\Gamma \left[
1+\frac{\epsilon }{2}\right] }{1-\frac{\epsilon }{2}}\left(
\frac{\bar{\chi}%
}{4\pi \mu ^{2}}\right) ^{1-\frac{\epsilon }{2}}
\end{equation}

\begin{equation}
M_{f}^{2}=\mu ^{\epsilon }\int \frac{d^{d}p}{\left( 2\pi \right) ^{d}}\;2\pi
\;f\left( p\right) \;\delta \left( p^{2}+\bar{\chi}\right)
\end{equation}
The gap equation becomes

\begin{equation}
\frac{1}{\lambda _{B}}\left\{ \bar{\chi}-M_{B}^{2}\right\} -\frac{\hbar
}{2}%
\mu ^{-\epsilon }\left[ M_{V}^{2}+M_{f}^{2}\right] =0  \label{logap2}
\end{equation}

We are now confronted with the formal need to show that Eq. (\ref{logap2})
admits finite solutions for $\bar{\chi}$ when $\epsilon \rightarrow 0$ \cite
{HK02}, as well as the physical need to show that the theory is reasonably
stable against changes in the distribution functions $f\left( p\right) $
\cite{EOS93,OS94,OSB96,OSS02}. Let us interpret this equation as defining $%
\bar{\chi}$ as a function of $M_{f}^{2}.$ Taking one derivative, we obtain

\begin{equation}
\left[ \frac{1}{\lambda _{B}}+\mu ^{-\epsilon }\frac{\hbar \Gamma \left[ 1+%
\frac{\epsilon }{2}\right] }{16\pi ^{2}\epsilon }\left( \frac{\bar{\chi}}{%
4\pi \mu ^{2}}\right) ^{-\frac{\epsilon }{2}}\right] \frac{d\bar{\chi}}{%
dM_{f}^{2}}-\frac{\hbar }{2}\mu ^{-\epsilon }=0
\end{equation}
This suggests defining a background field and renormalization scale
dependent effective coupling constant $\lambda $ from

\begin{equation}
\frac{d\bar{\chi}}{dM_{f}^{2}}\left( \bar{\chi},\mu ^{2}\right) =\frac{\hbar
}{2}\lambda \left( \bar{\chi},\mu ^{2}\right) \mu ^{-\epsilon }
\end{equation}
In other words

\begin{equation}
\frac{1}{\lambda _{B}}+\frac{\mu ^{-\epsilon }\hbar \Gamma \left[ 1+\frac{%
\epsilon }{2}\right] }{16\pi ^{2}\epsilon }\left( \frac{\bar{\chi}}{4\pi \mu
^{2}}\right) ^{-\frac{\epsilon }{2}}=\frac{1}{\lambda }  \label{renorcon}
\end{equation}

Now the gap equation

\begin{equation}
\frac{-1}{\lambda _{B}}M_{B}^{2}+\frac{1}{1-\frac{\epsilon }{2}}\left[
\frac{%
1}{\lambda }+\frac{\epsilon }{2\lambda _{B}}\right] \bar{\chi}-\frac{\hbar
}{%
2}\mu ^{-\epsilon }M_{f}^{2}=0
\end{equation}
shows that $\bar{\chi}=0$ when $M_{f}^{2}=M_{crit}^{2},$ given by

\begin{equation}
\mu ^{-\epsilon }\frac{\hbar }{2}M_{crit}^{2}=\frac{1}{\lambda
_{B}}M_{B}^{2}
\end{equation}
So assuming $M_{crit}^{2}$ to be finite, we may rewrite the gap equation as

\begin{equation}
\frac{1}{1-\frac{\epsilon }{2}}\left[ \frac{1}{\lambda }+\frac{\epsilon }{%
2\lambda _{B}}\right] \bar{\chi}-\mu ^{-\epsilon }\frac{\hbar }{2}\left[
M_{f}^{2}-M_{crit}^{2}\right] =0
\end{equation}
We see that it is possible to find a solution with finite propagators and $%
\bar{\chi}$ for any distribution function $f$, provided $M_{f}^{2}$ is
finite. On the other hand, (weak) thermalization would require that
any solution eventually converges to the thermal form Eq. (\ref{eqpdf}).
Therefore, we conclude that the $LO$ system does not thermalize.

\subsection{The $NLO$ approximation}

Since the $LO$ approximation admits a plurality of translation invariant
solutions, it is necessary to go at least to $NLO$ to study the issue of
thermalization. There is only one $NLO$ graph, consisting of a single $G$
loop and a single $H$ line (see Fig. 1). This graph leads to

\begin{equation}
\Gamma _{Q}^{NLO}=\left( -i\hbar \right) \left( \frac{-1}{2}\right) \left(
\frac{-N}{2Z_{B}\hbar }\right) ^{2}2N\left( \frac{\hbar }{N}\right)
^{3}c_{abc}c_{def}\int d^{d}xd^{d}y\;H^{ad}\left( x,y\right) G^{be}\left(
x,y\right) G^{cf}\left( x,y\right)
\end{equation}
Therefore, from

\begin{equation}
P_{ab}\left( x,y\right) =\frac{2}{\hbar }\frac{\delta \Gamma _{Q}}{\delta
G^{ab}(x,y)};\qquad Q_{ab}\left( x,y\right) =\frac{2}{\hbar }\frac{\delta
\Gamma _{Q}}{\delta H^{ab}(x,y)}
\end{equation}
we get

\begin{equation}
P_{ab}\left( x,y\right) =\frac{i\hbar }{Z_{B}^{2}}c_{acd}c_{bef}H^{ce}\left(
x,y\right) G^{df}\left( x,y\right)  \label{eq75}
\end{equation}

\begin{equation}
Q_{ab}\left( x,y\right) =\frac{i\hbar }{2Z_{B}^{2}}c_{acd}c_{bef}G^{ce}%
\left( x,y\right) G^{df}\left( x,y\right)
\end{equation}

Since $Q_{ab}$ does not depend on $H^{ab},$ we may solve the corresponding
equation

\begin{equation}
\frac{1}{\lambda _{B}}c_{ab}\delta (x,y)-i\left[ H^{-1}\right] _{ab}\left(
x,y\right) +Q_{ab}\left( x,y\right) =0
\end{equation}
in closed form. First, let us Fourier transform

\begin{equation}
\left[ H^{-1}\right] _{ab}\left( p\right) =\left( -i\right) \left(
\begin{array}{cc}
\frac{1}{\lambda _{B}}+Q_{11} & Q_{12} \\
Q_{21} & \frac{-1}{\lambda _{B}}+Q_{22}
\end{array}
\right)
\end{equation}
To lowest order in $1/N$, we may use the $LO$ propagators to compute the $%
Q^{\prime }s$. In particular, we get

\begin{equation}
Q_{11}=\mu ^{-\epsilon }\left[ Q_{V11}+Q_{f11}^{\left( 1\right)
}+Q_{f11}^{\left( 2\right) }\right]
\end{equation}
where

\begin{equation}
Q_{V11}\left( p\right) =\left( \frac{-i\hbar }{2}\right) \mu ^{\epsilon
}\int \frac{d^{d}q}{\left( 2\pi \right) ^{d}}\;\frac{1}{\left( p-q\right)
^{2}+\bar{\chi}-i\varepsilon }\frac{1}{q^{2}+\bar{\chi}-i\varepsilon }
\end{equation}

\begin{equation}
Q_{f11}^{\left( 1\right) }=\hbar \mu ^{\epsilon }\int \frac{d^{d}q}{\left(
2\pi \right) ^{d}}\;\frac{2\pi \;f\left( q\right) \;\delta \left(
q^{2}+\bar{%
\chi}\right) }{\left( p-q\right) ^{2}+\bar{\chi}-i\varepsilon }
\end{equation}

\begin{equation}
Q_{f11}^{\left( 2\right) }=\frac{i\hbar }{2}\mu ^{\epsilon }\int
\frac{d^{d}q%
}{\left( 2\pi \right) ^{d}}\;4\pi ^{2}\;f\left( q\right) \;\delta \left(
q^{2}+\bar{\chi}\right) \;f\left( p-q\right) \;\delta \left( \left(
p-q\right) ^{2}+\bar{\chi}\right)
\end{equation}
$\;$ We assume that $Q_{f11}^{\left( 1\right) }$ and $Q_{f11}^{\left(
2\right) }$ are well defined, and compute

\begin{equation}
Q_{V11}\left( p\right) =\frac{\hbar }{16\pi ^{2}}\frac{\Gamma \left[
1+\frac{%
\epsilon }{2}\right] }{\epsilon }\int_{0}^{1}dx\;\left( \frac{x\left(
1-x\right) p^{2}+\bar{\chi}}{4\pi \mu ^{2}}\right) ^{-\epsilon /2}
\end{equation}

Recalling the renormalization condition Eq. (\ref{renorcon}), we find

\begin{eqnarray}
\frac{1}{\lambda _{B}}+\mu ^{-\epsilon }Q_{V11} &=&\frac{1}{\lambda
}+\frac{%
\hbar \Gamma \left[ 1+\frac{\epsilon }{2}\right] }{16\pi ^{2}\epsilon }%
\left( \frac{\bar{\chi}}{4\pi }\right) ^{-\frac{\epsilon }{2}%
}\int_{0}^{1}dx\;\left[ \left( 1+x\left( 1-x\right) \left(
\frac{p^{2}}{\bar{%
\chi}}\right) \right) ^{-\epsilon /2}-1\right]  \nonumber \\
&\equiv &\frac{1}{\lambda }+Q_{Vren11}
\end{eqnarray}
$Q_{Vren11}$ is explicitly finite. Let us call $Q_{ren11}=Q_{Vren11}+\mu
^{-\epsilon }\left[ Q_{f11}^{\left( 1\right) }+Q_{f11}^{\left( 2\right) }%
\right] $. Since $Q_{22}=-Q_{11}^{\ast },$ we may also write

\begin{equation}
\frac{-1}{\lambda _{B}}+Q_{22}=\frac{-1}{\lambda }+Q_{ren22}
\end{equation}
where $Q_{ren22}=-Q_{Vren11}^{\ast }+\mu ^{-\epsilon }\left[
-Q_{f11}^{\left( 1\right) \ast }+Q_{f11}^{\left( 2\right) }\right] $.
Introduce $Q_{ret}=Q_{ren11}+Q_{12}$, $Q_{adv}=Q_{ren11}+Q_{21}=Q_{ret}^{%
\ast }$ and write this as (recall that $Q_{ren22}+Q_{ren11}=Q_{22}+Q_{11}=-%
\left( Q_{12}+Q_{21}\right) $)

\begin{equation}
\left[ H^{-1}\right] _{ab}\left( p\right) =\left( -i\right) \left(
\begin{array}{cc}
\frac{1}{\lambda }+Q_{ret}-Q_{12} & Q_{12} \\
Q_{21} & -\left( \frac{1}{\lambda }+Q_{adv}+Q_{12}\right)
\end{array}
\right)
\end{equation}
Call

\begin{eqnarray}
\Sigma ^{-1}\left( p\right) &=&\mathrm{\det }\left[ H^{-1}\right]
_{ab}\left( p\right) =\left[ \frac{1}{\lambda }+Q_{ret}-Q_{12}\right] \left[
\frac{1}{\lambda }+Q_{adv}+Q_{12}\right] +Q_{12}Q_{21}  \nonumber \\
&=&\left[ \frac{1}{\lambda }+Q_{ret}\right] \left[ \frac{1}{\lambda
}+Q_{adv}%
\right] +Q_{12}\left[ Q_{ret}-Q_{adv}\right] +Q_{12}\left[ Q_{21}-Q_{12}%
\right]  \nonumber \\
&=&\left[ \frac{1}{\lambda }+Q_{ret}\right] \left[ \frac{1}{\lambda
}+Q_{adv}%
\right]
\end{eqnarray}
Observe that $\Sigma \left( p\right) $ is real, finite and positive
definite. Then

\begin{equation}
H^{ab}\left( p\right) =i\Sigma \left( p\right) \left(
\begin{array}{cc}
\frac{1}{\lambda }+Q_{adv}+Q_{12} & Q_{12} \\
Q_{21} & -\left( \frac{1}{\lambda }+Q_{ret}-Q_{12}\right)
\end{array}
\right)
\end{equation}

\subsection{The $NLO$ density of states}

Beyond this point, our analysis will not depend upon the details of the
$NLO$
approximation, but only on a few structural features. One of these features
is the fact that the $NLO$ density of states is nonvanishing for $-p^{2}>9%
\bar{\chi}$. To establish this fact, it is enough to look at $P_{21}(p)$.
Recall that, from Eq. (\ref{p21}), we know that $P_{21}(p)$ is imaginary.
Its odd part determines the kernel $\mathbf{P}_{odd},$ and its even part the
noise kernel $\mathbf{N}$. $\mathbf{P}_{odd}$ determines $\gamma \left(
p\right) $ through Eq. (\ref{gamma}). For an actual translation invariant
solution, $\mathbf{N}$ and $\gamma $ are related by the fluctuation
dissipation relation. The density of states $\Delta $ and $\gamma $ are
related through Eq. (\ref{gamma2delta}). It is clear from this equation that
their zeroes are exactly the same, and so we only need to show that $\gamma
$
is non vanishing.

At $NLO$

\begin{eqnarray}
P_{21}\left( p\right) &=&\frac{-i\hbar }{Z_{B}^{2}}\int \frac{d^{d}q}{\left(
2\pi \right) ^{d}}\;H^{21}\left( q\right) G^{21}\left( p-q\right)  \nonumber
\\
&=&\frac{\hbar }{Z_{B}^{2}}\int \frac{d^{d}q}{\left( 2\pi \right) ^{d}}%
\;\Sigma \left( q\right) Q_{21}\left( q\right) G^{21}\left( p-q\right)
\nonumber \\
&=&\frac{-i\hbar ^{2}}{2Z_{B}^{4}}\int \frac{d^{d}q}{\left( 2\pi \right)
^{d}%
}\frac{d^{d}r}{\left( 2\pi \right) ^{d}}\;\Sigma \left( q\right)
G^{21}\left( r\right) G^{21}\left( q-r\right) G^{21}\left( p-q\right)
\end{eqnarray}
To simplify the analysis, we can make the rather drastic approximation $%
\Sigma \left( q\right) \sim \lambda ^{2},$ and use the $LO$ propagators.

\begin{eqnarray}
P_{21}\left( p\right) &=&\frac{-i\hbar ^{2}\lambda ^{2}}{2Z_{B}^{4}}\int
\frac{d^{d}q\Delta \left( q\right) }{\left( 2\pi \right) ^{d-1}}\frac{%
d^{d}r\Delta \left( r\right) }{\left( 2\pi \right) ^{d-1}}\frac{d^{d}s\Delta
\left( s\right) }{\left( 2\pi \right) ^{d-1}}\;\delta \left( q+r+s-p\right)
\nonumber \\
&&\left[ \theta \left( q^{0}\right) +f\left( q\right) \right] \left[ \theta
\left( r^{0}\right) +f\left( r\right) \right] \left[ \theta \left(
s^{0}\right) +f\left( s\right) \right]
\end{eqnarray}
It is convenient to write the integrals in terms of future oriented momenta
only. We get

\begin{equation}
P_{21}\left( p\right) =\frac{-i\hbar ^{2}\lambda ^{2}}{2Z_{B}^{4}}\left[
I_{3}+3I_{2}+3I_{1}+I_{0}\right]
\end{equation}
where

\begin{equation}
I_{3}=\int DqDrDs\;\delta \left( q+r+s-p\right) \left[ 1+f\left( q\right) %
\right] \left[ 1+f\left( r\right) \right] \left[ 1+f\left( s\right) \right]
\end{equation}

\begin{equation}
I_{2}=\int DqDrDs\;\delta \left( q+r-s-p\right) \left[ 1+f\left( q\right) %
\right] \left[ 1+f\left( r\right) \right] f\left( s\right)
\end{equation}

\begin{equation}
I_{1}=\int DqDrDs\;\delta \left( s-q-r-p\right) \left[ 1+f\left( s\right) %
\right] f\left( r\right) f\left( q\right)
\end{equation}

\begin{equation}
I_{0}=\int DqDrDs\;\delta \left( q+r+s+p\right) f\left( q\right) f\left(
r\right) f\left( s\right)
\end{equation}
and we have defined

\begin{equation}
Dq=\frac{d^{d}q}{\left( 2\pi \right) ^{d-1}}\;\theta \left( q^{0}\right)
\Delta \left( q\right)  \label{measure}
\end{equation}
It is clear that each of these integrals is nonnegative, so we only must
show that for arbitrary $p$ at least one is nonzero.

Let us assume the $LO$ density of states within the integrand, so that
momenta $q$, $r$ and $s$ are on-shell. Assume $-p^{2}>9\bar{\chi}$. If $%
p^{0}>0$, $I_{3}$ and $I_{2}$ are nonvanishing, while $I_{1}$ and $I_{2}$
are zero. If $p^{0}<0,$ it is the other way round. If $p^{0}>0$, moreover,
$%
I_{3}\left( p\right) >I_{0}\left( -p\right) $ and $I_{2}\left( p\right)
>I_{1}\left( -p\right) $, and so both the odd and even parts of $P_{12}$ are
nonvanishing, as we wanted to show. The fact that they are not only nonzero
but actually proportional to each other only obtains for a special form of
$%
f\left( p\right) ,$ indeed, a thermal form. We shall show this in next
Section.

It must be observed that in going from $P_{12}$ to $\mathbf{P}_{odd}$ there
is an extra factor of $1/N$ involved (cfr. Eq. (\ref{p12})), and so the
off-shell density of states, while non zero, is of higher order in $1/N$.

\section{The only translation invariant solutions to $NLO$ are thermal}

We may now show that the only translation invariant solutions of the $NLO$
equations are thermal. The solutions must satisfy the identity Eq. (\ref
{neccond}), which becomes

\begin{equation}
\int \frac{d^{d}q}{\left( 2\pi \right) ^{d}}\;\left[ H^{12}\left( q\right)
G^{12}\left( p-q\right) G^{21}\left( p\right) -H^{21}\left( q\right)
G^{21}\left( p-q\right) G^{12}\left( p\right) \right] =0
\end{equation}
>From the explicit solution for $H$

\begin{equation}
\int \frac{d^{d}q}{\left( 2\pi \right) ^{d}}\;\Sigma \left( q\right) \left[
Q_{12}\left( q\right) G^{12}\left( p-q\right) G^{21}\left( p\right)
-Q_{21}\left( q\right) G^{21}\left( p-q\right) G^{12}\left( p\right) \right]
=0
\end{equation}
Finally, use the $NLO$ approximation to $Q_{ab}$

\begin{equation}
\int \frac{d^{d}q}{\left( 2\pi \right) ^{d}}\frac{d^{d}r}{\left( 2\pi
\right) ^{d}}\;\Sigma \left( q\right) \left[ G^{12}\left( r\right)
G^{12}\left( q-r\right) G^{12}\left( p-q\right) G^{21}\left( p\right)
-G^{21}\left( r\right) G^{21}\left( q-r\right) \left( q\right) G^{21}\left(
p-q\right) G^{12}\left( p\right) \right] =0
\end{equation}
It is more usual to write this as

\begin{eqnarray}
0 &=&\left( 2\pi \right) ^{d}\int \frac{d^{d}q}{\left( 2\pi \right) ^{d}}%
\frac{d^{d}r}{\left( 2\pi \right) ^{d}}\frac{d^{d}s}{\left( 2\pi \right)
^{d}%
}\;\Sigma \left( p-q\right) \delta \left( q+r+s-p\right) \left\{
G^{12}\left( q\right) G^{12}\left( r\right) G^{12}\left( s\right)
G^{21}\left( p\right) \right.  \nonumber \\
&&\left. -G^{21}\left( q\right) G^{21}\left( r\right) \left( q\right)
G^{21}\left( s\right) G^{12}\left( p\right) \right\}
\end{eqnarray}
We recognize the usual Boltzmann collision term, with $\Sigma \left(
p-q\right) $ playing the role of cross section. Thus the only solutions must
be thermal.

There is one important observation to be made. Since this term is of order
$%
1/N$, in a strict power expansion we would use the $LO$ density of states
Eq. (\ref{lodos}) in the propagators. In the Introduction, we described this
as making a ''true'' approximation . In practice, it means that we would put
all momenta on mass - shell. Then only binary collisions would be possible,
and the Boltzmann term would admit solutions with nonvanishing chemical
potential. These do not exist in the exact theory, and so the ''true''
approximation does not describe thermalization.

We see, however, that if we keep the $NLO$ density of states the problem
disappears. In the Introduction, we called this procedure the \ ''exact''
way. Because the density of states is nonzero everywhere, it is possible for
one on - shell particle to decay into three off - shell ones, or vice versa,
for one off-shell particle with momentum $-p^{2}>9\bar{\chi}$ to decay onto
three on-shell particles (to show that this possibility is indeed open we
have shown explicitly in the last Section that the $NLO$ density of states
is nonvanishing in this region). Particle number is no longer conserved, and
only zero chemical potential is allowed.

Thermalization is described, but only as a higher than $NLO$
phenomenon, since it depends on $NLO$ corrections to the density
of states within an expression which is itself a $NLO$ construct.
We must consider if some of the $NNLO$ terms we have left out may
not bear on this process at a similar level. We shall see below
that this is indeed the case.

\section{Relaxation of the chemical potential}

We have seen so far that if nonperturbative corrections to the density of
states are allowed, then the only translation invariant solutions to the
Schwinger-Dyson equations are thermal propagators with vanishing chemical
potentials. These are also the only true equilibrium solutions of the full
theory. However, since we have stepped beyond the strict $NLO$
approximation, we must ask ourselves if terms we have discarded would not
affect thermalization at a similar level. To investigate this question, we
shall consider the relaxation of the chemical potential.

To place the issue in the simplest possible context, we shall assume that
only long wavelengths are involved, and adopt the Kadanoff-Baym approach.
That is, we shall assume that all two-point functions $G\left( x,y\right) $
may be written as functions of a relative variable $u=x-y$ and a center of
mass variable $X=\left( x+y\right) /2$, and that the $X$ dependence is weak,
loosely meaning that $\partial _{X}G\sim \left( 1/N\right) \partial _{u}G.$

We may introduce a mixed representation $G\left( X,p\right) $ by performing
a Fourier transform on the $u$ variable (cfr. Eq. (\ref{fourier})). For
fixed $X$, the manipulations in Section III are still valid, only now both
the density of states and the distribution function display an extra $X$
dependence (we refer the reader to ref. \cite{CHR00} and references therein
for a detailed discussion).

To obtain the dynamics of the distribution function $f\left( X,p\right) ,$
write the Schwinger-Dyson Eq. (\ref{schdy}) for $G^{21}\left( x,y\right) $
as

\begin{equation}
\frac{1}{Z_{B}}D\left( x,y\right) G^{21}\left( x,y\right) -\frac{1}{N}\int
d^{d}z\;\left[ P_{21}\left( x,z\right) G^{11}\left( z,y\right) +P_{22}\left(
x,z\right) G^{21}\left( z,y\right) \right] =0
\end{equation}
Recall that $D\left( x,y\right) =\left[ \partial^{2} _{x}-\bar{\chi}\left(
x\right) \right] \delta (x,y)$ and Fourier transform with respect to the
relative variable in each case to get, up to $1/N$ terms

\begin{equation}
\left[ -\Omega +iL\right] G^{21}\left( X,p\right) =\frac{Z_{B}}{N}\;\left[
P_{21}\left( X,p\right) G^{11}\left( X,p\right) +P_{22}\left( X,p\right)
G^{21}\left( X,p\right) \right]
\end{equation}
where $\Omega =p^{2}+\bar{\chi}\left( X\right) ,$ and $L$ is the Vlasov
operator

\begin{equation}
L=p\frac{\partial }{\partial X}-\frac{1}{2}\frac{\partial \bar{\chi}}{%
\partial X}\frac{\partial }{\partial p}
\end{equation}
Separating the imaginary part, we get the Boltzmann equation

\begin{equation}
Lf=I_{col}
\end{equation}
where (recall eqs. (\ref{p11}) to (\ref{p22}))

\begin{equation}
I_{col}=\frac{-iZ_{B}}{2N}\;\left[ P_{21}\left( X,p\right) G^{12}\left(
X,p\right) -P_{12}\left( X,p\right) G^{21}\left( X,p\right) \right]
\end{equation}

Observe that different approximations yield different collision integrals.
Of course, the condition $I_{col}=0$ for a translation invariant solution is
just the necessary condition Eq. (\ref{neccond}).

Manipulating the collision term as in the last Section, we get, to
$NLO$

\begin{eqnarray}
I_{col} &=&\frac{\left( 2\pi \right) ^{d}}{4N}\frac{\lambda ^{2}\hbar
^{2}}{%
Z_{B}^{3}}\int \frac{d^{d}q}{\left( 2\pi \right) ^{d}}\frac{d^{d}r}{\left(
2\pi \right) ^{d}}\frac{d^{d}s}{\left( 2\pi \right) ^{d}}\;\delta \left(
q+r+s-p\right)  \nonumber \\
&&\left\{ G^{12}\left( q\right) G^{12}\left( r\right) G^{12}\left( s\right)
G^{21}\left( p\right) -G^{21}\left( q\right) G^{21}\left( r\right) \left(
q\right) G^{21}\left( s\right) G^{12}\left( p\right) \right\}
\end{eqnarray}
where we leave the $X$ dependence implicit, and have approximated $\Sigma
\sim \lambda ^{2}$ for simplicity.

Since we are only interested in the relaxation of the chemical potential, we
may linearize the Boltzmann equation around equilibrium. Write the
distribution function as

\begin{equation}
f\left( X,p\right) =f_{0}\left( p\right) \left[ 1+\left( 1+f_{0}\right)
\delta f\left( X,p\right) \right]
\end{equation}
where $f_{0}$ is a thermal distribution with vanishing chemical potential.
Neglecting the variation in the density of states, which gives a higher
order contribution, we get

\begin{equation}
G^{21,12}=2\pi \Delta \left( p\right) \left[ \theta \left( \pm p^{0}\right)
+f_{0}+f_{0}\left( 1+f_{0}\right) \delta f\left( X,p\right) \right]
\end{equation}

A nonzero chemical potential corresponds to a $p$-independent perturbation
$%
\delta f\left( X,p\right) =\delta \mu \left( X\right) $. To isolate the
dynamics of the chemical potential, we may integrate both sides of the
Vlasov-Boltzmann equation with respect to $p.$ Since the distribution
function is even, there is no loss in resticting the integration region to
$%
p^{0}>0.$ We also write the collision integral in terms of integrals over
positive energy momenta, to get

\begin{equation}
B\frac{\partial }{\partial t}\delta \mu \left( X\right) =\frac{\left( 2\pi
\right) ^{d}}{2N}\frac{\hbar ^{2}}{Z_{B}^{3}}J  \label{relax}
\end{equation}
where (recall the momentum space measure Eq. (\ref{measure}))

\begin{equation}
B=\int Dp\;f_{0}\left( 1+f_{0}\right) p^{0}
\end{equation}

\begin{eqnarray}
J &=&\int DpDqDrDs\;\delta \left( q-r-s-p\right)  \nonumber \\
&&\left\{ f\left( q\right) \left[ 1+f\left( r\right) \right] \left[
1+f\left( s\right) \right] \left[ 1+f\left( p\right) \right] -\left[
1+f\left( q\right) \right] f\left( r\right) f\left( s\right) f\left(
t\right) \right\}
\end{eqnarray}

Linearizing the $J$ integral we get

\begin{equation}
J=-2K\delta \mu \left( X\right)
\end{equation}

\begin{equation}
K=\int DpDqDrDs\;\delta \left( q-r-s-p\right) \left[ 1+f\left( q\right) %
\right] \left[ 1+f\left( r\right) \right] \left[ 1+f\left( s\right) \right]
\left[ 1+f\left( p\right) \right] e^{-\beta q}
\end{equation}
The integral is nonzero, provided we use the $NLO$ density of states. If we
used the $LO\;$ density of states, then the argument of the delta function
would never vanish. This means we predict chemical potential relaxation on a
time scale which grows at least as $N^{2},$ since to the $1/N$ factor in Eq.
(\ref{relax}) we must add at least one more $1/N$ factor coming from the
off-shell density of states.

The problem is that there are terms, coming from the $NNLO$ approximation to
the $2PI$ $CTP$ $EA,$ that modify the collision term in such a way to
contribute to chemical potential relaxation at the same accuracy. Therefore
the $NLO$ prediction for the relaxation time is not accurate. We shall
conclude this Section by showing this explicitly.

\subsection{The $NNLO$ approximation and particle number violation}

As we already remarked, the power of $N$ associated to a given vacuum bubble
is the number of $G$ loops minus the number of $H$ lines. Therefore, in the
$%
NNLO$ approximation we must look for graphs with one more $H$ line than $G$
loops. However, a $2PI$ graph with $3$ $G$ loops must have no less than $5$
$%
H$ lines, so the allowed number of $G$ loops is only $1$ or $2.$ There are
only two graphs with $1$ $G$ loop and $2$ $H$ lines, but only one of them is
$2PI.$ (see Fig. 2) Similarly, there is only one $2PI$ graph with $2$ $G$
loops and $3$ $H $ lines (see Fig. 3). We therefore have two new graphs
contributing to the $2PI$ $CTP$ $EA.$

Variation of these graphs with respect to $G$ yields two new contributions
to the $P_{ab}$ (Figs. 4 and 5)

Let us consider Fig. 4, and replace the $H$ lines by their expansion in
powers of $\lambda$ (Fig. 6) The lowest order contribution yields the
setting sun diagram (Fig. 7).

Introducing the first correction to one of the $H$ lines gives the graph in
Fig. 8. Correcting both $H$ lines gives the graph in Fig. 9.

The second new graph in the $2PI$ $CTP$ $EA$ (Fig. 3) yields, upon
variation, the graph in Fig. 5.

Replacing the $H$ lines by their lowest order expression, we find another
contribution of the form of Fig. 8. The next order yields contributions
proportional to Figs. 9 and 10.

It has been shown in ref. (\cite{CHR00}) that the graphs in Fig.
9 and 10, when translated in terms of the collision integral,
describe scattering of $2 $ into $4$ particles and vice versa.
These scattering processes do not conserve particle number and
therefore contribute to the relaxation of the chemical potential
(we have shown explicitly the linearized collision operator in
ref. (\cite{CHR00})). The resulting contribution is at least of
the same order of magnitude as that found by allowing a
nonperturbative density of states in the $NLO$ collision term.

Therefore we conclude that the $NLO$ prediction for the relaxation time of
the chemical potential is not accurate. It is nevertheless remarkable that
the $NLO$ succeeds in predicting relaxation, in agreement with the claims of
refs. \cite{BC01} and \cite{B02}.

\section{How does macroscopic irreversibility appear?}

As we mentioned at the beginning of this paper, the issue of thermalization
in relativistic quantum fields, besides its concrete application to high
energy plasmas, is relevant to the larger question of the origin of the
thermodynamic arrow of time in physics. The $LO$ Large $N$ approximation
does not break time reversal invariance (it does not lead to thermalization
either) and one could jump to the conclusion that this is a general feature
of the Large $N$ perturbative scheme. However, as we have seen, the $NLO$
or, at worst, the $NNLO$ approximations show thermalization. It behooves us
to identify at which point time-reversal invariance has been broken, and
how.

To identify the crucial assumptions, let us return to the form Eq. (\ref
{classact}) of the action for the theory. A variation with respect to the
field yields the Heisenberg equations of motion for the field operator

\begin{equation}
\partial ^{2}\varphi ^{a}\left( x\right) -\chi \left( x\right) \varphi
^{a}\left( x\right) =0  \label{hei1}
\end{equation}
Multiplying from the left (say) by $\varphi ^{a}(y)$, taking expectation
values and summing over $a,$ we obtain (recall eqs. (\ref{eq12}) and (\ref
{eq20}))

\begin{equation}
\hbar \partial _{x}^{2}G^{21}(x,y)-\left\langle \chi \left( x\right) \varphi
^{a}\left( x\right) \varphi ^{a}(y)\right\rangle =0  \label{hei2}
\end{equation}

This Schwinger-Dyson equation is, of course, not a closed equation for the
propagator; it only relates the propagator to a higher correlation function.
If we wish to say something about this new correlation, one possibility is
to repeat the argument. We may begin from the Heisenberg equation for the
auxiliary field

\begin{equation}
\chi \left( x\right) =M_{B}^{2}+\frac{\lambda _{B}}{2}\varphi ^{a}\left(
x\right) \varphi ^{a}(x)
\end{equation}
and multiply by $\varphi ^{a}\left( x\right) \varphi ^{a}(y)$ to get

\begin{equation}
\left\langle \chi \left( x\right) \varphi ^{a}\left( x\right) \varphi
^{a}(y)\right\rangle =M_{B}^{2}\hbar G^{21}\left( x,y\right) +\frac{\lambda
_{B}}{2}\left\langle \varphi ^{b}\left( x\right) \varphi ^{b}(x)\varphi
^{a}\left( x\right) \varphi ^{a}(y)\right\rangle  \label{why4}
\end{equation}
or else we go back to Eq. (\ref{hei1}) to get

\begin{equation}
\partial _{y}^{2}\left\langle \chi \left( x\right) \varphi ^{a}\left(
x\right) \varphi ^{a}(y)\right\rangle -\left\langle \chi \left( x\right)
\varphi ^{a}\left( x\right) \chi \left( y\right) \varphi ^{a}\left( y\right)
\right\rangle =0
\end{equation}
In either case, yet another higher correlation is involved.

Comparing Eq. (\ref{hei2}) to Eq. (\ref{schdy}), we see that

\begin{equation}
\frac{1}{\hbar }\left\langle \chi \left( x\right) \varphi ^{a}\left(
x\right) \varphi ^{a}(y)\right\rangle =\bar{\chi}\left( x\right)
G^{21}(x,y)-%
\frac{1}{N}\int d^{4}z\;\left[ P_{\,\;1}^{2}\left( x,z\right) G^{11}\left(
z,y\right) +P_{\,\;2}^{2}\left( x,z\right) G^{21}\left( z,y\right) \right]
\end{equation}
Using Eq. (\ref{eq24}) for $\bar{\chi}$, and comparing to Eq. (\ref{why4}),
we get

\begin{eqnarray}
\frac{\lambda _{B}}{2\hbar }\left\langle \varphi ^{b}\left( x\right) \varphi
^{b}(x)\varphi ^{a}\left( x\right) \varphi ^{a}(y)\right\rangle &=&\frac{%
\lambda _{B}\hbar }{2}G^{21}(x,x)G^{21}(x,y)  \nonumber \\
&&-\frac{1}{N}\int d^{4}z\;\left[ P_{\,\;1}^{2}\left( x,z\right)
G^{11}\left( z,y\right) +P_{\,\;2}^{2}\left( x,z\right) G^{21}\left(
z,y\right) \right]
\end{eqnarray}
To be more concrete, observe that if Wick's theorem held, then

\begin{equation}
\left\langle \varphi ^{b}\left( x\right) \varphi ^{b}(x)\varphi ^{a}\left(
x\right) \varphi ^{a}(y)\right\rangle \sim \hbar ^{2}\left( 1+\frac{2}{N}%
\right) G^{21}(x,x)G^{21}(x,y)  \label{corre1}
\end{equation}
This suggests writing

\begin{equation}
\left\langle \varphi ^{b}\left( x\right) \varphi ^{b}(x)\varphi ^{a}\left(
x\right) \varphi ^{a}(y)\right\rangle \equiv \hbar ^{2}\left[
G^{21}(x,x)G^{21}(x,y)+\frac{1}{N}C\left( x,y\right) \right]  \label{corre2}
\end{equation}

The $NLO$ approximation consists in closing the Schwinger-Dyson hierarchy by
writing the $P_{ab}$'s as in Eq. (\ref{eq75}), whereby

\begin{equation}
\frac{\lambda _{B}}{2}C\left( x,y\right) =-i\int d^{4}z\;\left[ H^{21}\left(
x,z\right) G^{21}\left( x,z\right) G^{11}\left( z,y\right) -H^{22}\left(
x,z\right) G^{22}\left( x,z\right) G^{21}\left( z,y\right) \right]
\label{molchaos}
\end{equation}
Observe that the integrand vanishes when $z^{0}\succeq x^{0},$ $y^{0}.$ At
this point, time reversal symmetry has been broken.

Let us investigate further the mechanism for breaking time symmetry. First
decompose the $H$ propagators in singular and regular parts

\begin{equation}
H^{ab}\left( x,y\right) =H\left( x\right) c^{ab}\delta \left( x-y\right)
+H_{reg}^{ab}\left( x,y\right)
\end{equation}
Then $C\left( x,y\right) $ is split into a reducible and an irreducible term

\begin{equation}
C\left( x,y\right) =\frac{-2i}{\lambda _{B}}H\left( x\right)
G^{21}(x,x)G^{21}(x,y)+C_{irr}\left( x,y\right)  \label{corre3}
\end{equation}
Time symmetry is broken because the irreducible term $C_{irr}\left(
x,y\right) $ vanishes in the distant past while remains non-zero in the far
future.

For completeness, let us observe that Eq. (\ref{eq23}) shows that $H\left(
x\right) =i\lambda _{B}+O\left( \lambda _{B}^{2}\right) ,$ and therefore Eq.
(\ref{corre3}) leads to Eq. (\ref{corre1}) to leading order.

In summary, the scheme works because at a crucial point some
higher correlation ($C$ in eq. (\ref{corre2})) is replaced by a
perturbative expansion in terms of propagators. The replacement
assumes that the irreducible part of the higher correlation
($C_{irr}$ in eq. (\ref{corre3})) vanishes in the distant past,
in effect enforcing a variant of Boltzmann's molecular chaos
condition.

In order to restore time reversal symmetry, we ought to treat $C\left(
x,y\right) $ as a dynamical variable in its own right, for example, by
defining a higher generating functional with a new non local source coupled
to four fields. We refer the reader to ref. \cite{CH00} for a fuller
discussion of this issue.\newline

\textbf{Acknowledgements}
We acknowledge discussions of NLO scheme with Emil Mottola and Paul Anderson.
EC is supported in part by CONICET, UBA, Fundaci%
\'{o}n Antorchas and ANPCyT. BLH is supported in part by NSF grant
PHY98-00967 and their collaboration is supported in part by NSF
grant INT95-09847.

\newpage
\begin{figure}
\includegraphics[height=4cm]{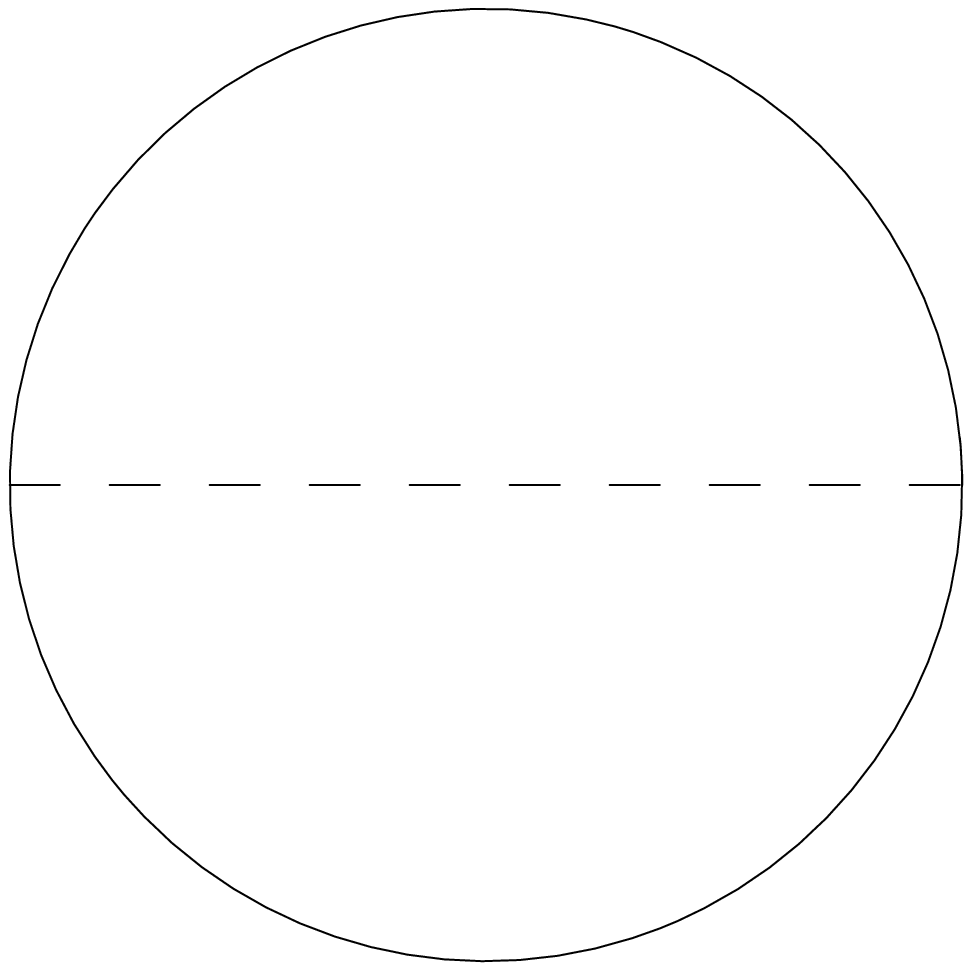}
\caption{Only $NLO$ contribution to the $2PI$ effective action. The full
line denotes a $G$ propagator, while the dotted line stands for $H$}
\end{figure}

\begin{figure}
\includegraphics[height=4cm]{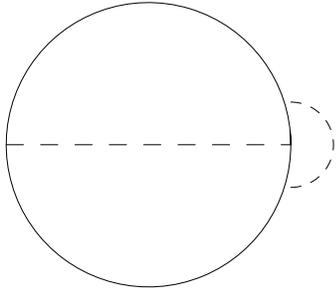}
\caption{Only $NNLO$ contribution to the $2PI$ effective action with a
single $G$ loop.}
\end{figure}

\begin{figure}
\includegraphics[height=4cm]{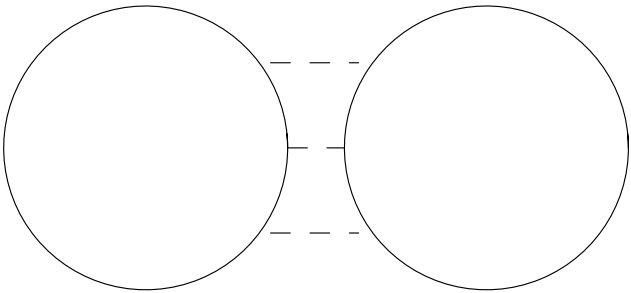}
\caption{Only $NNLO$ contribution to the $2PI$ effective action with $2$ $G$
loops.}
\end{figure}

\begin{figure}
\includegraphics[height=4cm]{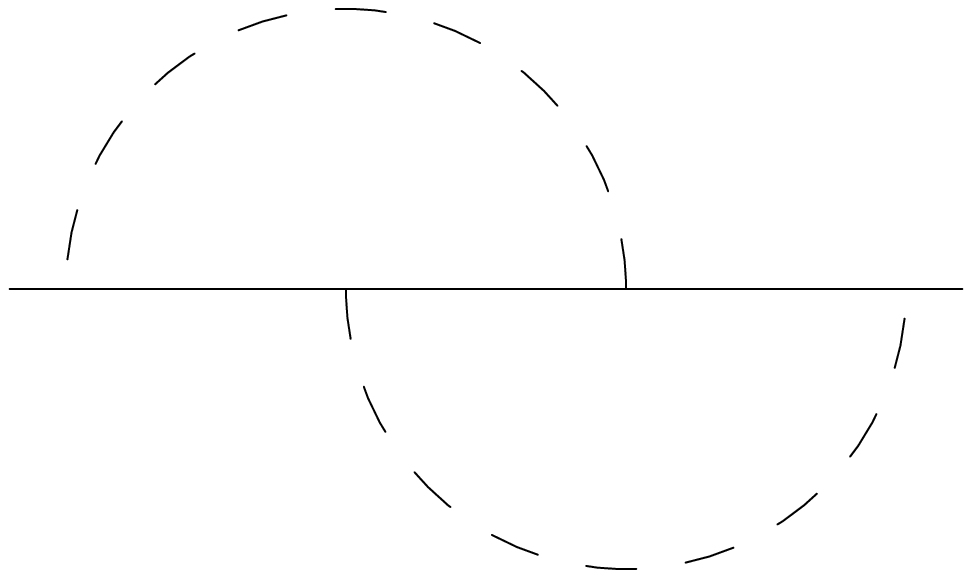}
\caption{Contribution to $P_{ab}$ from the variation of Fig. 2.}
\end{figure}

\begin{figure}
\includegraphics[height=4cm]{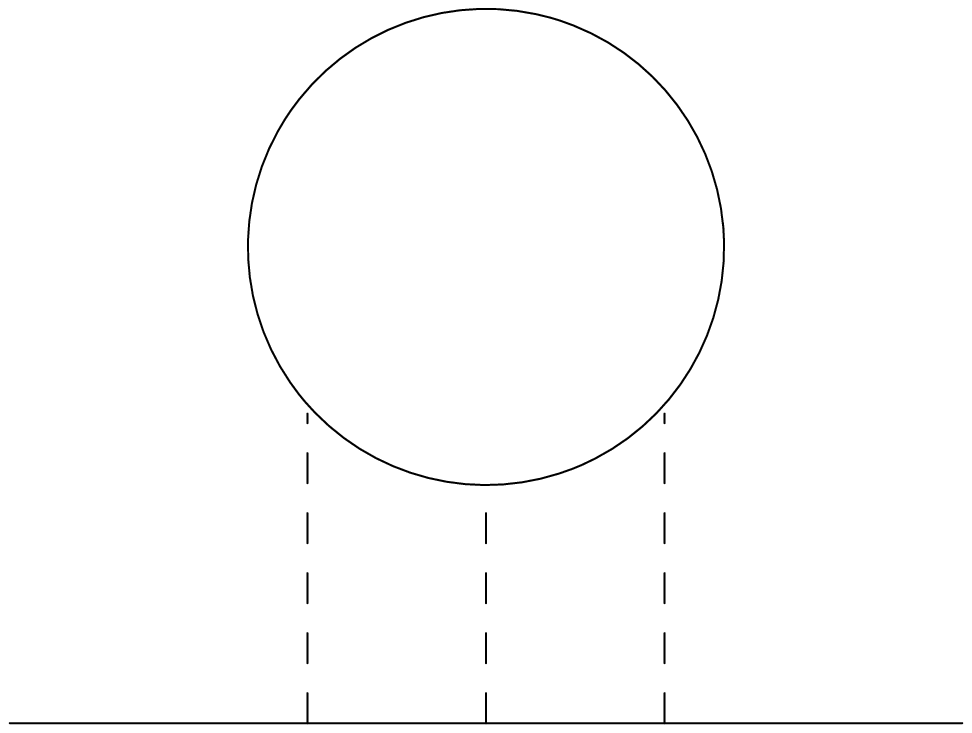}
\caption{Contribution to $P_{ab}$ from the variation of Fig. 3.}
\end{figure}

\begin{figure}
\includegraphics[height=4cm]{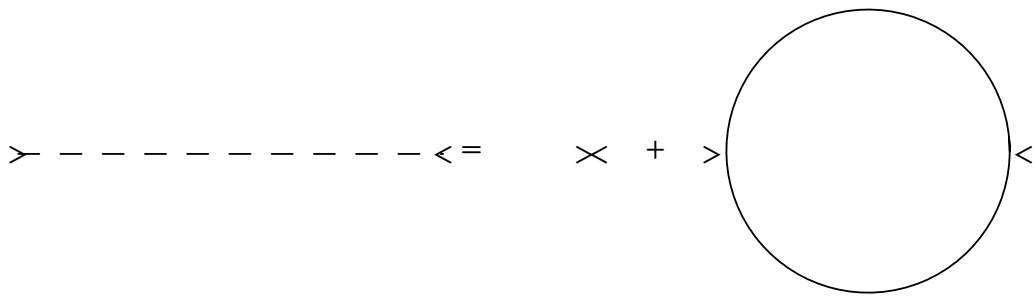}
\caption{Expansion of $H$ in powers of the coupling constant.}
\end{figure}

\begin{figure}
\includegraphics[height=4cm]{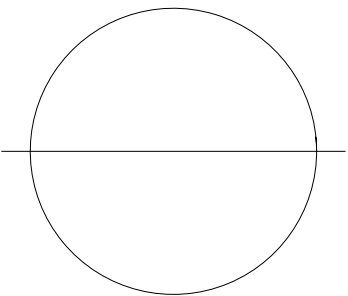}
\caption{Setting sun graph}
\end{figure}

\begin{figure}
\includegraphics[height=4cm]{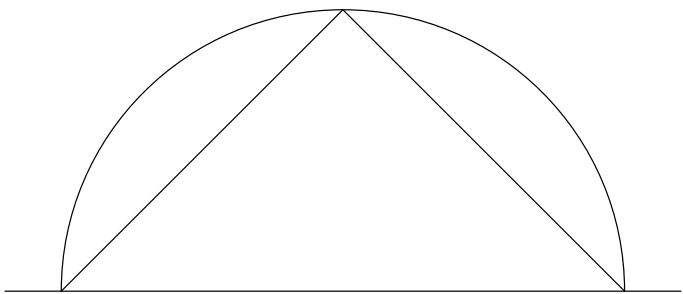}
\caption{First correction to Fig. 4}
\end{figure}

\begin{figure}
\includegraphics[height=4cm]{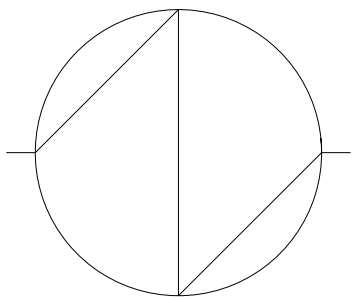}
\caption{Second correction to Fig. 4}
\end{figure}

\begin{figure}[b]
\includegraphics[height=4cm]{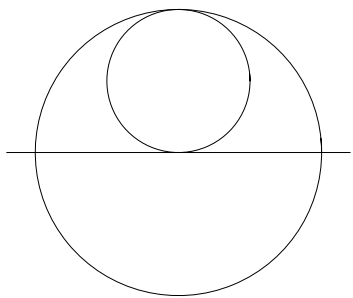}
\caption{Higher order correction to Fig. 5}
\end{figure}

\end{document}